\documentstyle[prb,aps,multicol,epsfig]{revtex}
\begin{document}
\draft
\title{Tunable decoherence in the vicinity of a normal
metal--superconducting junction}

\author{
Rodolphe Guyon$^{a}$,
Thierry Martin$^{a}$
and Gordey~B.\ Lesovik$^{a, b}$
}
\address{
$^{a}$Centre de Physique Th\'eorique,
Universit\'e de la M\'editerran\'ee,
Luminy Case 907, F--13288 Marseille Cedex 9, France}
\address{
$^{b}$L.~D.~Landau Institute for Theoretical Physics, 117940 
Moscow , Russia}
\maketitle
\begin{abstract}
The decoherence rate of a quantum dot coupled to a
fluctuating environment described by a  normal--metal
superconductor junction is considered. The density--density
correlator at low frequencies  constitutes the kernel which
enters the calculation of the phase coherence time.  The density
fluctuations are connected to the finite frequency
current--current correlations in the point contact via the
continuity equation. Below and above the gap, at zero 
temperature the density correlator contains spatial oscillations
at half of the Fermi wave length on the normal side. As the bias 
crosses the superconducting gap,
the opening of new scattering channels enhances  the decoherence
rate dramatically, suggesting the possibility of tuning the
decoherence rate in a controllable manner. 
\end{abstract}
\begin{multicols}{2}

\section{Introduction}

The problem of decoherence in mesoscopic systems has remained a
central issue for two decades. With the
possible advent of quantum computers, it now occupies an even
more important role,  as the limitations associated with various 
decoherence mechanisms provide the fundamental 
working limits of these devices. 
Proposals for studying decoherence in
Aharonov--Bohm (AB) type geometries have been made \cite{Stern}. 
Experiments in the last decade have
provided information on the  phase shift suffered by an
electron propagating through a quantum dot
\cite{Yacoby,Schuster}. These interference experiments show
that coherent propagation through a dot is indeed possible. 
A few years ago, decoherence was introduced
artificially  in these same AB devices by placing
a quantum point contact in the vicinity of the dot\cite{Buks}. 
Alternatively, dephasing has been studied experimentally 
using a phase sensitive double dot detector \cite{Sprinzak}.
Because quantum transport is a stochastic process, current noise
or charge fluctuations act as a dissipative environment
coupled to the dot, providing decoherence without particle 
transfer to this environment.      

In the present work, the decoherence rate is computed
for the situation where the discrete level (the dot)
is coupled electrostatically to the fluctuating charges 
of a normal metal--superconductor junction with an arbitrary 
bias. In the Andreev regime, at zero temperature, it is expected
that the  calculation of the decoherence rate is similar to that 
of a normal metal point contact except that the charge 
of the carriers is replaced by the Cooper pair
charge\cite{Kozhevnikov}.  This simple analogy  fails both at
finite temperatures and at voltage  biases superior to the
superconducting gap.  Specifically, in both the Andreev and the
sub--gap regime, the decoherence 
rate depends crucially on the range of the potential 
which couples the dot to its environment: when this range
is lowered below the Fermi wavelength, oscillatory terms in the
density--density correlator contribute substantially to the 
decoherence rate. 
Even more stunning is the behavior of the decoherence rate 
as the voltage bias crosses the gap. The opening of new 
scattering channels accounts for an additional charge noise,
and thus provokes a sharp increase in this rate.
In principle, this could allow to tune the system 
from a ``quantum'' behavior to a ``classical'' behavior 
in a controllable manner.  

The computation of the decoherence
rate in the present work, which typically involves the
calculation of a density--density correlator in the limit 
of zero frequency, uses
as a starting  point known results
for the finite frequency current noise
\cite{lesovik_martin_torres,torres_martin_lesovik}, rather than
a direct computation of the density fluctuations.
Indeed, the charge  fluctuations are directly connected to the
current
 fluctuations using the continuity equation, which holds
in 
 second quantized notation, as will be shown below. 

\begin{figure}
\epsfxsize 8 cm
\centerline{\epsffile{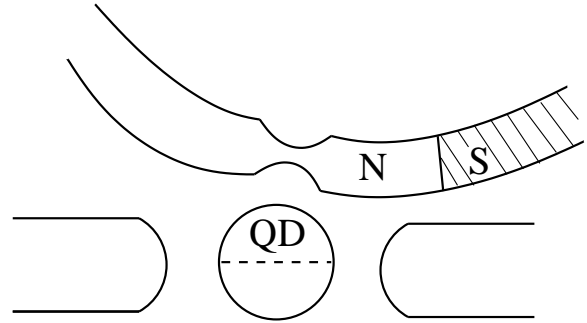}}
\medskip
\caption{\label{ns_dot}
A quantum dot (QD), which   contains a sharp level, is connected
to two semi--infinite leads (lower half of the figure).
In its vicinity, a fluctuating current flows through
a nanoscopic (single channel), normal metal--superconductor
junction (upper half). Broadening of the level occurs in the
presence of an electrostatic coupling between the dot and the
normal  side of the NS wire.}
 \end{figure} 
 
Density--density correlations have been recently computed on a
formal level \cite{Martin Buttiker} in the geometry of Fig.
\ref{ns_dot}. In contrast, here one is interested both in
ultra--small junctions and in the transition to the above--gap
regime. Moreover, it will be shown that decoherence also 
occurs for an ``ideal'' NS junction, the superconducting--normal 
metal analog of an adiabatic point contact. In all of the above, 
spatial oscillations of the density--density correlations are
shown to occur, and to enter the computation of the decoherence
rate.  To our knowledge these oscillations have not been 
described in previous work. 

The paper is organized as follows. The model with its basic
assumptions is described in section \ref{model}. The relationship
between the phase coherence time and the  density--density
fluctuations has been established by several authors
\cite{Levinson1,Aleiner,Levinson2} and is reviewed
briefly in Appendix \ref{app_decoherence_rate}.
 The main emphasis is put on the
analysis of the density--density correlations of the
normal--metal superconducting quantum dot, first in the
Andreev regime (section \ref{correlator andreev}).
Expressions which apply to the case where the disorder
potential is smooth -- a superconducting adiabatic point
contact -- are provided. Finally, the case of a bias which is 
superior to the gap is exposed in section 
\ref{decoherence rate above}, and its consequence on the
decoherence rate is illustrated numerically using the Blonder
Tinkham Klapwijk (BTK) model\cite{BTK}.

\section{Dephasing mechanism \\
and NS junction}
\label{model}

A (small) quantum dot, which for practical purposes here is
represented by a discrete, sharp energy level, is located in the
proximity of a (single channel) normal metal--superconductor (NS)
junction 
 (Fig. \ref{ns_dot}).
 The dot is connected to two
semi--infinite leads, and can in principle be part of an
interferometer such as the ones studied in Refs.\cite{Buks,Sprinzak}.
The dot is coupled by Coulomb forces to the fluctuating charges
located in a NS point
 contact (SPC). To be specific, it is
assumed  that the electrostatic coupling is restricted to the
normal side of the junction, as the dephasing mechanism is
expected to be more efficient when the bias voltage on the NS
wire is below the superconducting gap. 

The system is described by a Hamiltonian which characterized the
dot, the NS junction and the coupling between the two:
\begin{equation}
 H_C=c^\dagger c\int dx U(x)
\psi^\dagger(x)\psi(x)
 \end{equation}
where $\psi(x)$  and $c$ ($\psi^\dagger(x)$ 
and $c^\dagger$) are fermion creation and annihilation
operators in the NS junction and in the dot.
The potential $U(x)$ depends on the location 
of the charges in the NS junction. While it originates from
long range Coulomb forces, in practical situations it is
screened by the surrounding metallic gates. Later on, different 
ranges will be specified for $U(x)$ in order to observe their
consequence on the decoherence rate.     

The ``standard'' procedure\cite{Levinson1,Aleiner} for computing
the decoherence rate $\tau_\phi^{-1}$ is to identify an
exponential decay in time of the dot electron Greens function.
Even though there is no electron leak from the dot to the NS
junction, the level in the dot acquires a finite width due to
the coupling with the fluctuations in the junction. The dot
Greens function thus acquires a non--oscillatory component: 
\begin{equation} G(t)=\langle T[c(t)
c^\dagger(0)]\rangle_{N\!S}\propto e^{-t/\tau_\phi}
e^{-i\epsilon_0 t/\hbar} \label{dot green} 
\end{equation} 
where $\tau_\phi$ is the decoherence time, and the bracket notation 
$\langle~\rangle_{N\!S}$ implies that an average over
the NS environment has been taken. The following result
for the decoherence rate neglects the back--effect of the dot 
on the NS junction:
\begin{equation} 
{1\over \tau_\phi}={1\over
\hbar^2}\int_{-\infty}^{+\infty} dt
K(t) \label{coherence rate}
\end{equation}
with the Kernel defined as:
\begin{eqnarray}
K(t)&=&{1\over 2}\int dx_1\int dx_2 
U(x_1)U(x_2)\nonumber\\
&~&~~~~\times \langle\langle \rho(x_1,t)\rho(x_2,0)+
\rho(x_2,0)\rho(x_1,t)\rangle\rangle
\label{kernel is density}
\end{eqnarray}
where the notation $\langle\langle~ \rangle\rangle$ means that
the average densities have been subtracted. 

The main issue to compute this rate is to specify the 
density--density correlator using the property that 
electrons and holes on the normal side, and possibly
quasi--particles on the superconducting side, are scattered at the
NS junction.  The scattering matrix specifies the amplitudes
of the transmitted/reflected particles. It allows to give the
asymptotic behavior of   the electron and hole wave functions
 away from the
junction \cite{lesovik_martin_torres,de Jong,Anantram}. 
Here, the notations of previous work 
\cite{torres_martin_lesovik}  
are used for convenience. 

On the normal side of the NS point contact, 
the electron and hole wave functions associated
with a particle $\alpha=e,h$ which originates from 
side $j=N,S$ are given by: 
\begin{eqnarray}
u_{j,\alpha} (x,t) &\simeq& 
{\delta_{1j}\delta_{\alpha e}\over (hv_+)^{1/2}} 
(e^{i k_{+}x}+s_{N\!jee}e^{-i k_{+}x})\nonumber\\
&~&~~~+ {\delta_{\alpha h}\over (hv_+)^{1/2}}
s_{N\!jeh}e^{-i k_{+}x}
\\ 
v_{j,\alpha} (x,t) &\simeq&
{\delta_{1j}\delta_{\alpha h}\over (hv_-)^{1/2}}
(e^{-i k_{-}x}+s_{N\!jhh}e^{i k_{-}x})+
\nonumber\\ 
&~&~~~+ {\delta_{\alpha e}\over (hv_-)^{1/2}} 
s_{N\!jhe} e^{i k_{-}x} 
\end{eqnarray}
where the electron and hole momenta are specified by
$k_\pm=\sqrt{2m \left( \mu_S \pm \epsilon \right)}/\hbar$, with
$\mu_S$ the chemical potential of the superconductor, 
which assumed to be large compared to both the gap $\Delta$ 
and the bias $eV$.
$s_{ij\alpha\beta}$ is the amplitude for a particle 
of type $\beta=e,h$ which is incident from side $j$ to be
scattered as a particle of type $\alpha=e,h$ in 
reservoir $i$. When discussing the Andreev regime, 
both electrons and holes are incident only from the 
normal side, and the indices $i,j=N$ will therefore
be dropped in sections \ref{current and density} and
 \ref{correlator andreev}, but will be restored in section
 \ref{decoherence rate above}. 

Finally, the electron/hole distribution
function on the normal side are given by:
$f_{N\!e,h}=\{1+\exp[(\epsilon \mp eV)/k_B\Theta]\}^{-1}$
while on the superconducting side, incident 
holes and electrons have the same distribution 
function, with $V=0$.

\section{Current and density fluctuations}
\label{current and density}

As described before\cite{lesovik_martin_torres}, the
statistical average of the current--current operator is obtained
by performing the Bogolubov transformation \cite{de Gennes} on
the current operator. In the past, it was for
the most part computed at equal locations in connection with the
current noise across the NS junction. For the decoherence time,
it is necessary to keep the two  locations $x_1$ and $x_2$
separate. However, as the current correlators in Refs.
\cite{lesovik_martin_torres,torres_martin_lesovik}
are expressed with separate time arguments $t_1$ and $t_2$, 
the generalization to separate spatial arguments 
is straightforward: 
\begin{eqnarray} 
 \nonumber && \langle \langle
I(x_1,t_1)I(x_2,t_2)
 \rangle \rangle = \frac{e^2 \hbar^2}{2m^2
}\sum_{\tilde{\alpha},\tilde{\beta}}
 \int_0^{+\infty}
\int_0^{+\infty} 
 d\varepsilon d\varepsilon' \Big\{
\hspace{1.8cm}
\nonumber
\\
&& \hspace{0.5cm}
~  f_{\tilde{\alpha}}(\epsilon)(1-f_{\tilde{\beta}}(\epsilon'))
e^{i (\varepsilon'- \varepsilon)(t_2-t_1)/\hbar}\nonumber
\\ 
&&\hspace{1.0cm} \times 
[(u_{\tilde{\beta}}\stackrel{\leftrightarrow}{\partial} 
\!\! u_{\tilde{\alpha}}^{*})_{t_1,x_{1}}+
(v_{\tilde{\beta}}\stackrel{\leftrightarrow}{\partial} 
\!\! v_{\tilde{\alpha}}^{*})_{t_1,x_{1}}\rbrack\nonumber
\\ 
&&\hspace{1.0cm} 
\times \lbrack(u_{\tilde{\beta}}^{*}\stackrel{\leftrightarrow}{\partial} 
\!\! u_{\tilde{\alpha}})_{t_2,x_{2}}+
(v_{\tilde{\beta}}^{*}\stackrel{\leftrightarrow}{\partial} 
\!\! v_{\tilde{\alpha}})_{t_2,x_{2}}]
 \nonumber\\  && \hspace{0.5cm} +
f_{\tilde{\alpha}}(\epsilon) f_{\tilde{\beta}}(\epsilon') e^{-i (\varepsilon +
\varepsilon')(t_2-t_1)/\hbar}\nonumber\\ 
&& \hspace{1.0cm} \times 
 (u_{\tilde{\beta}}^* \!\! \stackrel{\leftrightarrow}{\partial} 
\!\! v_{\tilde{\alpha}}^*)_{t_1,x_1} 
[
(u_{\tilde{\alpha}} \!\! \stackrel{\leftrightarrow}{\partial} 
\!\! v_{\tilde{\beta}})_{t_2,x_2} + (u_{\tilde{\beta}} \!\!
\stackrel{\leftrightarrow}{\partial} \!\! v_{\tilde{\alpha}})_{t_2,x_2}
] \nonumber\\
&& \hspace{0.5cm}
+ (1-f_{\tilde{\alpha}}(\epsilon))(1-f_{\tilde{\beta}}(\epsilon'))
e^{i (\varepsilon + \varepsilon')(t_2-t_1)/\hbar}\nonumber
\\ 
&& \hspace{1cm} \times
 [(u_{\tilde{\alpha}} \!\! \stackrel{\leftrightarrow}{\partial} 
\!\! v_{\tilde{\beta}})_{t_1,x_1} + (u_{\tilde{\beta}} \!\!
\stackrel{\leftrightarrow}{\partial} \!\! v_{\tilde{\alpha}})_{t_1,x_1}
] (u_{\tilde{\beta}}^* \!\! \stackrel{\leftrightarrow}{\partial}
\!\! v_{\tilde{\alpha}}^*)_{t_2,x_2}  \Bigr\}, \nonumber 
\end{eqnarray}
\begin{equation}
\label{general current correlator}
\end{equation}

where $\tilde{\alpha}=(\alpha,i)$ is a short hand notation
combining  the reservoir from which the particle, electron or
hole, is incident. Eq. (\ref{general current correlator})
constitutes the starting point for computing both finite
frequency noise and the zero frequency noise in the presence of
a  local harmonic  perturbation, such as in the Non Stationary
AB effect in NS junctions\cite{lesovik_martin_torres} which was 
recently detected experimentally\cite{Kozhevnikov}.
Here it constitutes the starting point for the computation of the
density--density correlator, and it is valid both in the Andreev
regime and above gap, provided that the proper distribution
functions are specified on the superconducting side. 

The continuity equation allows to relate the current
operator to the density operator as it holds in second
quantized form:

\begin{equation}
\rho(x,\omega)=
{1\over i\omega}
\overrightarrow{\nabla}.\overrightarrow{J}(x,\omega)
\end{equation}

This allows to write a connection formula between the
nonlocal current noise correlator and the density--density
correlator at finite frequency:   
\begin{eqnarray}
\langle \langle \rho(x_1,\omega)\rho(x_2,-\omega) \rangle \rangle
&=&  \int_{-\infty}^{+\infty} {dt\over \omega^2}
e^{i\omega t}\nonumber\\ &~&~
\times\partial_{x_{1}}\partial_{x_{2}}\langle \langle I(0)I(t)
\rangle \rangle  
\end{eqnarray}
Here the $\omega=0$ density--density correlations will be  needed.
Taking the derivative with respect to the positions,
the $\omega^2$ term in the denominator is canceled, for 
all bias regimes, giving a finite contribution to the density
fluctuations.

\section{Density correlator in \\
the Andreev regime}
\label{correlator andreev}

In the Andreev regime, 
there are only two types of particles (electrons
and holes), so the current--current correlator at finite
temperatures and bias takes the form: 
\begin{eqnarray}
\nonumber
&&
\langle \langle I(x_1,t_1)I(x_2,t_2) \rangle \rangle_A =
\frac{e^2 \hbar^2}{2m^2 }
\int_0^{+\infty} \int_0^{+\infty} 
d\varepsilon d\varepsilon' \Big\{
\hspace{1.8cm}
\\
\nonumber
&& \hspace{0.5cm}
  f_{e}(\epsilon)(1-f_{h}(\epsilon'))
e^{i (\varepsilon'- \varepsilon)(t_2-t_1)/\hbar}
\\ \nonumber
 && \hspace{1.0cm} \times
[(u_{h}\stackrel{\leftrightarrow}{\partial} 
\!\! u_{e}^{*})_{t_1,x_{1}}+
(v_{h}\stackrel{\leftrightarrow}{\partial} 
\!\! v_{e}^{*})_{t_1,x_{1}}\rbrack
\\ \nonumber
&&\hspace{1.0cm} \times
\lbrack(u_{h}^{*}\stackrel{\leftrightarrow}{\partial} 
\!\! u_{e})_{t_2,x_{2}}+
(v_{h}^{*}\stackrel{\leftrightarrow}{\partial} 
\!\! v_{e})_{t_2,x_{2}}]
 \\ \nonumber && \hspace{0.5cm} +
f_{h}(\epsilon)(1-f_{e}(\epsilon'))
e^{i (\varepsilon'- \varepsilon)(t_2-t_1)/\hbar}
\\ \nonumber
 && \hspace{1.0cm} \times
[(u_{e}\stackrel{\leftrightarrow}{\partial} 
\!\! u_{h}^{*})_{t_1,x_{1}}+
(v_{e}\stackrel{\leftrightarrow}{\partial} 
\!\! v_{h}^{*})_{t_1,x_{1}}\rbrack
\\ \nonumber
&&\hspace{1.0cm} \times
\lbrack(u_{e}^{*}\stackrel{\leftrightarrow}{\partial} 
\!\! u_{h})_{t_2,x_{2}}+
(v_{e}^{*}\stackrel{\leftrightarrow}{\partial} 
\!\! v_{h})_{t_2,x_{2}}] \\ \nonumber && \hspace{0.5cm} +
f_{e}(\epsilon) f_{h}(\epsilon') e^{-i (\varepsilon +
\varepsilon')(t_2-t_1)/\hbar}
\\ \nonumber
&&\hspace{1.0cm} \times
 (u_{h}^* \!\! \stackrel{\leftrightarrow}{\partial} 
\!\! v_{e}^*)_{t_1,x_1} 
 [
(u_{e} \!\! \stackrel{\leftrightarrow}{\partial} 
\!\! v_{h})_{t_2,x_2} + (u_{h} \!\!
\stackrel{\leftrightarrow}{\partial} \!\! v_{e})_{t_2,x_2}
] \\
\nonumber
&& \hspace{0.5cm}
+ f_{h}(\epsilon) f_{e}(\epsilon') e^{-i (\varepsilon +
\varepsilon')(t_2-t_1)/\hbar}
\\ \nonumber
&&\hspace{1.0cm} \times
 (u_{e}^* \!\! \stackrel{\leftrightarrow}{\partial} 
\!\! v_{h}^*)_{t_1,x_1} 
[
(u_{h} \!\! \stackrel{\leftrightarrow}{\partial} 
\!\! v_{e})_{t_2,x_2} + (u_{e} \!\!
\stackrel{\leftrightarrow}{\partial} \!\! v_{h})_{t_2,x_2}
] \\
\nonumber
&& \hspace{0.5cm}
+ (1-f_{e}(\epsilon))(1-f_{h}(\epsilon'))
e^{i (\varepsilon + \varepsilon')(t_2-t_1)/\hbar}
\\ \nonumber
&& \hspace{1cm}\times
 [(u_{e} \!\! \stackrel{\leftrightarrow}{\partial} 
\!\! v_{h})_{t_1,x_1} + (u_{h} \!\!
\stackrel{\leftrightarrow}{\partial} \!\! v_{e})_{t_1,x_1}
] (u_{h}^* \!\! \stackrel{\leftrightarrow}{\partial}
\!\! v_{e}^*)_{t_2,x_2} \\
\nonumber
&& \hspace{0.5cm}
+ (1-f_{h}(\epsilon))(1-f_{e}(\epsilon'))
e^{i (\varepsilon + \varepsilon')(t_2-t_1)/\hbar}
\\ 
&& \hspace{1cm}\times
 [(u_{h} \!\! \stackrel{\leftrightarrow}{\partial} 
\!\! v_{e})_{t_1,x_1} + (u_{e} \!\!
\stackrel{\leftrightarrow}{\partial} \!\! v_{h})_{t_1,x_1}
] (u_{e}^* \!\! \stackrel{\leftrightarrow}{\partial}
\!\! v_{h}^*)_{t_2,x_2} \Bigr\}
\nonumber 
\end{eqnarray} 
\begin{equation}
\label{current current}
\end{equation}

To proceed, one makes use of the unitarity of the 
scattering matrix, combined with the
time reversal symmetry of electrons and holes
in the Andreev regime (energy dependence of scattering
coefficients neglected): $s_{he}^{*}=- s_{eh}$ and  
$s_{ee}^{*}= s_{hh}$. 
Taking the Fourier transform of the current--current 
correlator, the following relation between 
the incoming and the scattered wave numbers 
is obtained:
\begin{eqnarray} 
k_{\pm}^{2}-k_{\pm}'^{2} &=& \pm
\frac{2m}{\hbar}\omega
\label{wave numbers}\end{eqnarray}
Moreover, the standard
simplifications are made ($\mu_S\gg \omega$):
once the spatial derivatives are taken and the 
$\omega^2$ proportionality in 
$\partial_{x_{1}}\partial_{x_{2}}\langle \langle
I(x_1,t_{1})I(x_2,t_{2}) \rangle \rangle$ is identified, 
the wave vectors of electrons and holes are assumed to
be  equal to $k_F$. 

The density correlator in the Andreev regime becomes:
\begin{eqnarray}
\nonumber
&&
\langle \langle \rho(x_1,\omega)\rho(x_2,-\omega) \rangle
\rangle_A = \frac{e^2 }{2 \pi^2 \hbar v_F^2}
 \int_0^{+\infty} d\epsilon  \Big\{
\hspace{1.8cm}
\\
\nonumber
&& \hspace{0.2cm}
f_e(\epsilon) (1-f_h(\epsilon -\omega))\Theta(\epsilon-\omega)
\\
\nonumber
&& \hspace{0.5cm} 
\times 4 \vert s_{eh}\vert^{2}\lbrack \vert
s_{ee}\vert^{2}+s_{ee}^{*}e^{2i
k_{F}x_{2}}+s_{ee}e^{-2i k_{F}x_{1}}+e^{2i
k_{F}(x_{2}-x_{1})}\rbrack
\\
\nonumber
&& \hspace{0.2cm}
+f_h(\epsilon) (1-f_e(\epsilon -\omega))\Theta(\epsilon-\omega)
\\
\nonumber
&& \hspace{0.5cm} 
\times 4 \vert s_{eh}\vert^{2}\lbrack \vert
s_{ee}\vert^{2}+s_{ee}^{*}e^{2i
k_{F}x_{1}}+s_{ee}e^{-2i k_{F}x_{2}}+e^{2i
k_{F}(x_{1}-x_{2})}\rbrack
\\ \nonumber
&& \hspace{0.2cm}
- f_{e}(\epsilon) f_{h}(-\epsilon +
\omega)
\Theta(-\epsilon+\omega)  \\ \nonumber && \hspace{0.5cm}
\times \vert s_{eh}\vert^2
 \lbrace 1 - (\vert s_{eh}\vert ^{2} - s_{ee} s_{hh}) + 
s_{ee} e^{-2 i k_{F} x_2 } + s_{hh} e^{2 i k_{F} x_2 } \rbrace
\\ \nonumber
&& \hspace{0.2cm}
+ f_{h}(\epsilon) f_{e}(-\epsilon+\omega)
\Theta(-\epsilon+\omega)
 \\ \nonumber && \hspace{0.5cm}
\times \Big[ 2 \vert s_{ee}\vert ^{2} \lbrace 
\vert s_{ee}\vert ^2 + 1 +
s_{ee}(e^{-2 i k_F x_1 } + 1/ 2 e^{-2 i k_F x_2 })
\\ \nonumber && \hspace{0.8cm}
+ s_{ee}^{*} ({e^{2 i k_F x_1 }+1/2  e^{2 i k_F x_2 }})
+cos(2 k_F(x_2-x_1))\rbrace 
\nonumber
 \\
&& \hspace{0.8cm} + s_{ee}^{*} e^{2 i k_{F} x_2 }(1 + s_{ee}^{*} e^{2 i k_{F}
x_1} ) 
\nonumber
 \\
&& \hspace{0.8cm}
+ s_{ee} e^{-2 i k_{F} x_2 }(1 + s_{ee} e^{-2 i k_{F} x_1} ) \Big]\Big\}
\nonumber \\
\label{rho rho distributions}
\end{eqnarray}

Further assuming that the temperature $k_B\Theta<\Delta$,
and neglecting the energy dependence of the scattering matrix
coefficients, the thermal integrations are performed in 
Appendix \ref{appendix finite temp}.

At zero temperature ($\omega>0$) the only interval which survives
is the one  specified by the Fermi functions $f_e(\epsilon)
(1-f_h(\epsilon-\omega))$ (first term in Eq. (\ref{rho rho
distributions})).  At low frequencies, and $k_B\Theta=0$, the
double integral in energy is $eV$ and the correlator becomes:
\begin{eqnarray}
&&\langle \langle \rho(x_1,\omega)\rho(x_2,-\omega) \rangle
\rangle_{A,\omega =0} = \frac{2 e^3 V \vert s_{eh}\vert
^2}{\pi^2 \hbar v_F^2}\nonumber\\ && \hspace{1cm} \times
\lbrack \vert s_{ee}\vert^{2}+s_{ee}^{*}e^{2ik_{F}x_{2}}
+s_{ee}e^{-2i k_{F}x_{1}}+e^{2ik_{F}(x_{2}-x_{1})}\rbrack
\nonumber\\
\label{rho rho zero}
\end{eqnarray}

Note the remarkable fact that the low frequency
density--density correlator has an oscillatory spatial
dependence with wavelength $\lambda_F/2$. These oscillatory 
terms give a significant contribution to the kernel
of Eq. (\ref{kernel is density}) when the envelope function 
$U(x)$ is short ranged.
Together with the above approximations, the decoherence 
rate in the Andreev regime can be expressed in terms of the
Fourier components $\widetilde{U}(q)$ of the envelope potential
at $q=0$ and at $q=\pm 2k_F$: 
\begin{eqnarray}   
\frac{1}{\tau_\Phi^A} &=& \lim_{\omega\rightarrow 0}\frac{e^2}
{\hbar^3 v_F^2 } \int_0^{+\infty} 
d\varepsilon f_{e}(\epsilon)(1-f_{h}(\epsilon-\omega))
\Theta(\epsilon-\omega) 
\nonumber \\
&& \times
[s_{he}^*(s'_{hh}\widetilde{U}(0)+\widetilde{U}(-2 k_F))
\nonumber \\
&& \hspace{1cm}
- s'_{eh}(s^*_{he}\widetilde{U}(0)+\widetilde{U}(-2 k_F))]
\nonumber\\
&& \times  
[s_{he}(s^{'*}_{hh}\widetilde{U}(0)+\widetilde{U}(2 k_F))
\nonumber \\
&& \hspace{1cm}
- s^{'*}_{eh}(s_{ee}\widetilde{U}(0)+\widetilde{U}(2 k_F))] ,
\end{eqnarray}

where prime denotes quantities evaluated at $\epsilon - \omega$.

The case of a superconducting adiabatic point contact (SAPC) is
now considered momentarily, to highlight the different role
played by charge and current fluctuations for the decoherence
rate. This constitutes the NS analog of the adiabatic point
contact studied in Ref. \cite{glazman_lesovik}. This situation
assumes a sharp NS interface with perfect  Andreev reflection,
adjacent to a scalar potential  on the normal side which varies
slowly on the scale of  the Fermi wave length
.
While the 
Andreev reflection processes have a unit
probability,
it is
 necessary to take into account the 
dependence of the electron
 and hole wave function amplitude
on the normal side due to the smooth disorder potential
following a quasi--classical/WKB approximation. For simplicity,
here
 only results at zero temperature, in the Andreev regime
are
 presented.
 
Instead of redefining the electron and hole wave function in
this limit, it is more convenient to directly substitute the
expressions for the semi--classical matrix elements 
in the density correlator of Eq. (\ref{rho rho distributions}):
$\vert s_{he} \vert ^2 = 1$ and $ s_{ee} = s_{h\!h}
=0 $. The wave numbers of electrons and holes, which now have
acquired  a spatial dependence, still satisfy the relationships
of Eq. (\ref{wave numbers}).

For the SAPC, the only contribution which survives is the one 
which is directly proportional to the Andreev reflection
probability:
\begin{eqnarray}
&&\langle \langle \rho(x_1,\omega)\rho(x_2,-\omega)\rangle\rangle_{W\!K\!B}
=  
\nonumber
\\
&&
\frac{16 e^{2}}{h^2} 
\int\limits_{0}^{\infty} 
\,d\epsilon 
{f_{e}(\epsilon)(1-f_{h}(\epsilon - \omega))
\over \sqrt{v_{+} (x_1)v_{-}(x_2)}}\nonumber
\\
&& ~~~~~~~~~\times{\exp\left[ \int_{x_1}^{x_2}i (k'(x)+k(x))
dx\right]\over \sqrt{
(v'_{+}(x_1)+v_{+}(x_1)) (v'_{-}(x_2)+v_{-}(x_2))}}
\end{eqnarray}  

Specifying that the distribution functions 
are step functions, this simplifies to:

\begin{eqnarray}
&&\langle \langle \rho(x_1,\omega)\rho(x_2,-\omega)\rangle\rangle_{W\!K\!B}
= \frac{16 e^{3}V }{h^2 \sqrt{v_{+} (x_1)v_{-}(x_2)}}
\nonumber\\&& \hspace{0.5cm}
\times
\frac{\exp\left[ \int_{x_1}^{x_2}i (k'(x)+k(x))
dx\right]}
{\sqrt{(v'_{+}(x_1)+v_{+}(x_1))
(v'_{-}(x_2)+v_{-}(x_2))}}
\label{density density sapc}
\end{eqnarray}
Note the analogy of Eq. (\ref{density density sapc})
with the expressions derived for normal, adiabatic point 
contact\cite{lesovik}. This illustrates 
that even for a NS junction  with ideal transmission,
the decoherence rate does not vanish, as
substantial density fluctuations are present, although the 
current fluctuations are reduced due to the Pauli principle
(which operates on electrons and holes on the normal side).

\section{Decoherence rate at\\ 
arbitrary bias}
\label{decoherence rate above}

In the previous expressions, contributions 
where quasi--particles are transmitted into the superconductor
were discarded. Here,
the calculation of the density--density correlator proceeds 
as before, choosing for simplicity zero temperature, 
a constraint which excludes some combinations of 
$f_{Ne,h}$ and $f_{Se,h}$. The calculation proceeds 
in a similar way as in the Andreev limit, except that 
processes involving quasi--particle emission from the
superconductor now contribute.
Operating the simplifications on the wave vectors
as before, and using the continuity equation, the density 
correlator becomes:
\begin{eqnarray}
\nonumber
&&
\langle \langle \rho(x_1,\omega)\rho(x_2,-\omega) \rangle
\rangle 
- \langle \langle \rho(x_1,\omega)\rho(x_2,-\omega)
\rangle \rangle_A =\nonumber\\
&&\frac{2 e^2}{h^2 v_F^2 }
\int_0^{+\infty} 
d\varepsilon   \Big\{
\hspace{1.8cm}\nonumber
\\
&& \hspace{0.2cm} f_{Ne}(\epsilon)
(1-f_{Se}(\epsilon-\omega))
\Theta(\epsilon-\omega)\nonumber
\\
 && \hspace{0.5cm} \times
(s'_{N\!She}s^*_{N\!Nhe} + s'_{N\!See} (e^{-2 i k_F x_1} +
 s^*_{N\!Nee}))
\nonumber\\
&&\hspace{0.5cm} \times(s^{'*}_{N\!She}s_{N\!Nhe} +
 s^{'*}_{N\!See} (e^{2 i k_F
x_2} + s_{N\!Nee}))\nonumber \\  && \hspace{0.2cm} +f_{Ne}(\epsilon)
(1-f_{Sh}(\epsilon-\omega))
\Theta(\epsilon-\omega)\nonumber
\\
 && \hspace{0.5cm} \times
(s'_{N\!Shh} s^{*}_{N\!Nhe}- s'_{N\!Seh}(e^{-2 i k_F x_1} +
 s^*_{N\!Nee}))
\nonumber\\
&&\hspace{0.5cm} \times(s^{'*}_{N\!Shh} s_{N\!Nhe}-
 s^{'*}_{N\!Seh}(e^{2 i k_F x_2}
+ s_{N\!Nee}))\nonumber  \\  &&
\hspace{0.2cm} +f_{Ne}(\epsilon)
 f_{Se}(-\epsilon+\omega) 
\Theta(-\epsilon+\omega)\nonumber\\
 && \hspace{0.5cm} \times
s^{'*}_{N\!See} s^*_{N\!Nhe}
( s'_{N\!See} s_{N\!Nhe} + 
s_{N\!She}(e^{2 i k_F x_2} + s'_{N\!Nee})
)\nonumber
\\
&& \hspace{0.2cm} +f_{Ne}(\epsilon)
f_{Sh}(-\epsilon+\omega) 
\Theta(-\epsilon+\omega)\nonumber\\ 
 && \hspace{0.5cm} \times
s^{'*}_{N\!Seh} s^*_{N\!She}
( s'_{N\!Seh} s_{N\!Nhe} +
s_{N\!Shh}(e^{2 i k_F x_2} + s'_{N\!Nee})
) \Big\}\nonumber\\
\label{density_density}
\end{eqnarray}
where $\langle \langle \rho(x_1,\omega)\rho(x_2,-\omega) \rangle
\rangle_A$ is the density--density correlator with Andreev
scattering contributions only (yet the energy integral ranges
from $0$ to $eV$, above the gap). 

Taking into account that the decoherence rate involves only zero
frequency density--density correlations, and that quasi--particle energies
are always positive, the decoherence rate reads:
\begin{eqnarray}
\nonumber
\frac{1}{\tau_\Phi}&-&\frac{1}{\tau_\Phi^A} = \frac{e^2}{\hbar^3  v_F^2}
\int_0^{eV}  d\varepsilon   \Big\{
\hspace{1.8cm}\nonumber
\\ && \hspace{0.2cm}
(s_{N\!She}s^*_{N\!Nhe}\widetilde{U}(0)\nonumber
\\ && \hspace{1.0cm}
 + s_{N\!See} (\widetilde{U}(-2 k_F ) +
s^*_{N\!Nee}\widetilde{U}(0)))
\nonumber
\\ && \hspace{0.2cm}
\times (s^*_{N\!She}s_{N\!Nhe}\widetilde{U}(0) 
\nonumber
\\ && \hspace{1.0cm}
+
s^*_{N\!See} (\widetilde{U}(2 k_F ) + s_{N\!Nee}\widetilde{U}(0))) \nonumber
\\ 
&&  + 
(s_{N\!Shh} s^{*}_{N\!Nhe}\widetilde{U}(0)
\nonumber
\\ && \hspace{1.0cm}
- s_{N\!Seh}(\widetilde{U}(-2 k_F ) +
s^*_{N\!Nee}\widetilde{U}(0)))
\nonumber
\\ && \hspace{0.2cm}\times
 (s^*_{N\!Shh} s_{N\!Nhe}\widetilde{U}(0)-
\nonumber
\\ && \hspace{1.0cm}
s^*_{N\!Seh}(\widetilde{U}(2 k_F ) + s_{N\!Nee}\widetilde{U}(0)))\Big\}
\nonumber\\
\end{eqnarray}

In order to enquire about the effect of the oscillatory terms, 
a Gaussian potential profile with a specific width $\xi$ is 
chosen:
\begin{eqnarray} U(x) &=&\frac{U_0}{\xi}
e^{-\frac{x^2}{2 \xi^2}} \nonumber\\ \widetilde{U}(k) &=& 
\frac{U_0}{\sqrt{2 \pi}} e^{-\frac{k^2 \xi^2}{2}}~,
\end{eqnarray} 
where $\xi$ represents the screening length of the Coulomb
interaction due to the surrounding metallic gates. For
nanoscopic  dots and junctions, or alternatively for the large
wavelengths which apply to semiconductors--2D electron gas
structures, it is becoming conceivable that $\lambda_F$ could
become larger than $\xi$.

With this particular choice, 
the decoherence rate can be expressed as:
\begin{eqnarray}
\frac{1}{U_0^2}&& \left(\frac{1}{\tau_\Phi}-\frac{1}{\tau_\Phi^A
}\right)= \frac{e^2 }{2 \pi \hbar^3 v_F^2 } \int_0^{eV} 
d\varepsilon   \Big\{
\hspace{1.8cm}\nonumber
\\  &&  
(s_{N\!She}s^*_{N\!Nhe} + s_{N\!See} (e^{- 8 \pi^2 \frac{\xi^2}{\lambda_F^2}} +
s^*_{N\!Nee}))\nonumber 
\\ 
&&~~~~~ \times (s^*_{N\!She}s_{N\!Nhe}
+
s^*_{N\!See} (e^{- 8 \pi^2 \frac{\xi^2}{\lambda_F^2}} + s_{N\!Nee})) \nonumber
\\ 
&&  + 
(s_{N\!Shh} s^{*}_{N\!Nhe}- s_{N\!Seh}(e^{- 8 \pi^2 \frac{\xi^2}{\lambda_F^2}} +
s^*_{N\!Nee}))\nonumber \\ 
&&~~~~~ \times (s^*_{N\!Shh} s_{N\!Nhe}-
s^*_{N\!Seh}(e^{- 8 \pi^2 \frac{\xi^2}{\lambda_F^2}} + s_{N\!Nee}))\Big\}
\nonumber\\
\label{tau_above_range}\end{eqnarray}
where the Andreev decoherence rate reads:
\begin{eqnarray}
\nonumber
&& \frac{1}{\tau_\Phi^A U_O^2}=
\frac{e^2 }{2 \pi \hbar^3 v_F^2 } \int_0^{eV} 
d\varepsilon   \Big\{
\hspace{1.8cm}
\nonumber\\
&&~~~
[s_{N\!Nhe}^*(s_{N\!Nhh}+ e^{- 8 \pi^2 \frac{\xi^2}{\lambda_F^2}})
- s_{N\!Neh}(s^*_{N\!Nhe}+e^{- 8 \pi^2 \frac{\xi^2}{\lambda_F^2}})]
\nonumber\\
&&~ \times
[s_{N\!Nhe}(s^*_{N\!Nhh}+e^{- 8 \pi^2 \frac{\xi^2}{\lambda_F^2}})
- s^*_{N\!Neh}(s_{N\!Nee}+e^{- 8 \pi^2 \frac{\xi^2}{\lambda_F^2}})]
\Big\} \nonumber
\\
\label{tau_andreev_range}
\end{eqnarray}

The results are now illustrated by plotting the decoherence rate 
as a function of bias voltage, for different values of the
potential range $\xi$. Because first, it is relevant
to enquire about the role of disorder in the NS junction, and
second, the complete energy dependence of the scattering matrix
coefficients  is required, a model with a minimal  set of
parameters, the BTK model \cite{BTK}, is chosen. Expressions for
the scattering matrix elements of this model 
are known\cite{torres_martin_lesovik}:
there, the same model was chosen to
enquire about the singularities in the frequency dependent noise
in a NS junction, for arbitrary biases. Recall that the NS
junction is described by a delta function  barrier
$V(x)=V_B\delta(x)$ and a step--wise pair potential
$\Delta(x)=\Delta \Theta(x) $, both located at  the  normal
metal--superconductor interface. The strength of the (normal) 
barrier is represented by the variable $Z=mV_B/\hbar^2k_F$,
which for intermediate values between high and low
transparencies is of order $1$. 

\begin{figure} \epsfxsize 8 cm
\centerline{\epsffile{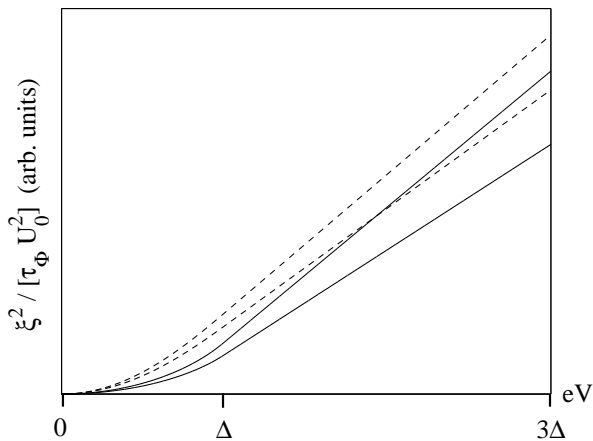}}
\medskip
\caption{\label{decoh_rate1}
Decoherence rate $\xi^2[\tau_\phi U_0^2]^{-1}$ 
as a function of bias voltage in the Andreev 
and sub--gap regime, 
for barrier transparencies: $Z=0.1$ (dashed line), at
$\xi=0.15\lambda_F$ (top) and at $\xi=10\lambda_F$
(bottom); $Z=1.0$ (full line) at $\xi=10\lambda_F$ (top)
and at $\xi=0.15\lambda_F$ (bottom).}  
\end{figure}  

For high to intermediate barrier transparencies, 
the decoherence rate is plotted in Fig. \ref{decoh_rate1}.
This rate is normalized to the interaction
strength squared $(U_0/\xi)^2$, which allows to plot 
the curves corresponding to $Z=0.1$ and $Z=1.0$
on the same scale. In addition,
results are illustrated for two values of the range $\xi$
for comparison: $\xi=10\lambda_F$ corresponds to a range 
where the negative exponentials in Eqs. 
(\ref{tau_above_range}) and (\ref{tau_andreev_range})
can be neglected; on the opposite, for $\xi=0.15\lambda_F$ 
these exponentials give a significant contribution.

Deep within the Andreev regime $eV\ll \Delta$, the decoherence
rate have a linear dependence on the bias (not shown), as the
scattering matrix coefficients are essentially energy
independent. 
The rates computed with these parameters
show no noticeable change in slope once the voltage bias crosses
the gap.  
This linear dependence is expected for large, sub--gap biases, 
because transport across the NS junction then becomes dominated
by single quasi--particle transfer: similar behavior was recently
observed for the finite frequency noise of NS junctions 
\cite{torres_martin_lesovik}. 

It is difficult to predict the behavior of 
$\tau_\Phi^{-1}$ when
the  energy dependence of the scattering matrix coefficients 
becomes noticeable, i.e. for biases near the
superconducting gap. In Fig. \ref{decoh_rate1}, no
qualitative changes of the different decoherence rates 
are observed until $eV\sim \Delta$, above which
the rates show a gradual crossover to the linear dependence 
on the bias. Yet, the results plotted in Fig. \ref{decoh_rate1}
show that the relative magnitude of these (normalized)
decoherence rates depends in a non trivial manner on the
potential range $\xi$: for intermediate barriers ($Z=1.0$)
the oscillatory terms in the density--density 
correlator (Eq. \ref{density_density}) tend to reduce
$\xi^2[\tau_\phi U_0^2]^{-1}$, while for high transparency
barriers, they increase the decoherence rate.  
In addition, for $eV\sim 1.9\Delta$ the curves for 
$\xi=10\lambda_F$ corresponding to intermediate and weak
barriers cross, and the rate corresponding to $Z=1.0$
is larger beyond this. 
\begin{figure}
\epsfxsize 8 cm
\centerline{\epsffile{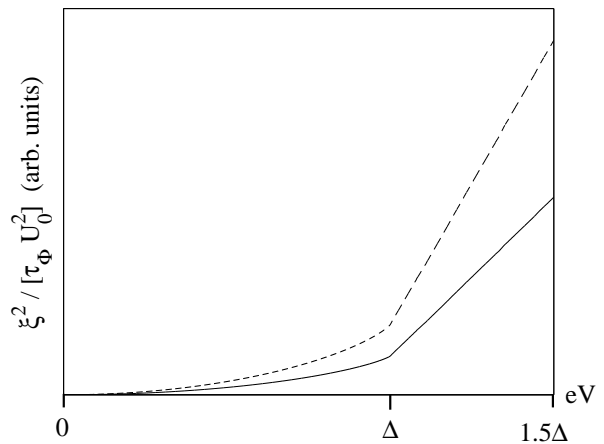}}
\medskip
\caption{\label{decoh_rate2}
Same as Fig.\ref{decoh_rate1} for opaque barriers
($Z=10.$): $\xi=10\lambda_F$ (dashed line),
and $\xi=0.15\lambda_F$ (full line).} 
\end{figure}  

Most interesting is the behavior of the decoherence rate
for opaque barriers ($Z=10.$), which is illustrated in 
Fig. \ref{decoh_rate2}. For both potential ranges 
$\xi=10\lambda_F$ and $\xi=0.15\lambda_F$, $\tau_\Phi^{-1}$
displays a clear cusp as the bias crosses the gap. 
This cusp corresponds to the opening of new scattering 
channels -- electron and hole quasi--particle transfer from the
normal metal to the superconductor -- which gave previously 
a gradual change -- and no obvious crossover region --
for higher transparencies  (Fig. \ref{decoh_rate1}): here, the
crossover region is confined in a very small interval with a
width lesser than a few percent of $\Delta$.
Above the gap, the linear dependence on the bias
is  recovered.

Note that the decoherence rate should in principle be highest
when the current undergoes strong temporal fluctuations:
one naively expects\cite{Aleiner} 
that it is related to the 
shot--noise fluctuations\cite{lesovik_martin_landauer_buttiker} 
$\sim T(1-T)$. 
Here, $T$ stands either for 
the Andreev reflection probability (below the gap)
or for the transmission probability of quasi--particles
(sub--gap regime). Indeed, the curves corresponding to 
$Z=10$ (Fig. \ref{decoh_rate2}) could not be plotted on the same
scale as  in Fig. \ref{decoh_rate1} because the corresponding
rate  is too small.

The drastic change depicted in Fig. \ref{decoh_rate2} could
possibly be  exploited to switch the decoherence rate from
``large'' to ``small'', thus allowing to control the degree of
coherence -- from quantum to ``classical'' (phase incoherent)
transport -- in
the neighboring quantum dot (Fig. \ref{ns_dot}).   In addition,
note that for existing NS junctions 
opaque barriers are likely to be the norm rather than the
exception, which renders an experimental check of this effect
plausible. 

\section{Conclusion}

The low frequency density--density correlations of a NS point
contact have been extracted from the finite frequency noise
characteristic, using the continuity equation. This density
correlator enters the  computation of the dephasing rate of an
isolated quantum system with a discrete spectrum.  

In the shot noise regime, the  density fluctuation function
contains terms which  oscillate with half a Fermi wavelength. 
Because this length
scale is considerably larger in semiconductors than in metals,
and given the ongoing progress in nano--fabrication techniques, it
is suggested  that the detection of such oscillations, or their
effects in the dephasing rate, could indeed be possible.
The role of these oscillatory terms is noticeable for 
weak disorder even in the (sub--gap) Andreev regime.
 
Expressions of the density--density correlator described in
section \ref{correlator andreev} for a NS
junction were generalized to the case of a
Superconducting Adiabatic Point Contact, using a 
semi--classical
scheme. Note that this implies that decoherence can even 
occur when the NS junction located near the quantum dot
has ideal charge transmission properties (pure Andreev scattering).

While the computation of the decoherence rate bears similarities 
with that of a normal--normal point contact \cite{lesovik}, 
no simple extension so far has been provided when both
Cooper pairs and quasi--particles are transmitted in the
superconductor. The competition between the two charge transfer
processes has been a central issue
in the present work. 

Regardless of the disorder strength, the decoherence rate
rises sharply as the voltage bias is increased above the 
superconducting gap. Yet for opaque barriers, this transition 
is even more dramatic, as indicated by the cusp obtained in  
Fig. \ref{decoh_rate2}. 
This crossover is due to the opening of additional
scattering channels, which renders the fluctuating environment 
more noisy. It is conceivable that this increase of
the decoherence rate could have some potential applications.

In particular consider a quantum bit (qubit), implemented by a 
quantum dot apparatus: the coupling to a neighboring NS junction
can render the system classical (besides the tunneling
processes in and out of the dot) or quantum in a perfectly
controllable manner. Decoherence, reduced abruptly by lowering 
the voltage bias below the gap, could be used to determine when
the quantum evolution of the qubit is supposed to start, or to
end. Other applications include the classical resetting of a
quantum computer. 

The present approach has dealt with a single channel NS wire. 
Clearly a generalization to a many
channel wire would bring this proposal closer to experimental
realizations, in carbon nanotubes and so on. 
Yet, the main goal here has been to demonstrate the variability 
of the decoherence rate on the bias voltage as it crosses the
gap. Because a specific model (BTK) was used to plot the
decoherence rate, the additional complexity of dealing with more
than one channel could have rendered these results less explicit,
but extensions are indeed possible.

Finally, for more
immediate experimental applications, 
 it is necessary to work
with superconducting materials which
 have a small enough gap
that imposing a voltage bias at the NS
 junction does not cause
substantial heating on the normal side.
 Indeed, the present
theoretical approach assumes that the 
 transport across the
junction is purely elastic, in the
 mesoscopic regime. Such
superconductors with gaps of the order 
 of hundreds of $mK$ are
readily available, and allow presently
 to observe a crossover
behavior in the current noise above and
 below the gap
\cite{private_glattli}.
 \acknowledgements

J. Torres is gratefully acknowledged for help with
the BTK scattering matrix elements. 

\appendix
\section{Calculation of the decoherence rate}
\label{app_decoherence_rate}

The starting point for the decoherence rate is the dot Green 
function:
\begin{equation}
G(t)=-i\langle T[c(t)c^\dagger(0)]\rangle
\end{equation}
Operators are in the Heisenberg picture.  Specifying the time
evolution:    \begin{eqnarray}
<c(t)&& c^\dagger(0)> = 
\nonumber \\ 
&&< e^{-i \epsilon_0 t/\hbar} T_t 
e^{-{i\over \hbar} \int_0^{t} dt' \int dx U(x)
\psi^\dagger(x,t)\psi(x,t)} c(0)c^\dagger(0)> \nonumber\\
\label{dot green2} \end{eqnarray}

Following Levinson \cite{Levinson1,Levinson2}, it is assumed
that while the NS point contact has an effect on the dot, the
reverse is not taken into account. The Greens function
acquires a non--oscillatory time dependence: \begin{equation}
<c(t) c^\dagger(0)> = e^{-i \epsilon_0 t} \exp[-\phi(t)]
\label{dot green3}
\end{equation}
where the decoupling between the dot and point contact degrees
of freedom allows to compute the average as in a Gaussian
process:  \begin{eqnarray}
\phi(t) &\simeq& -\ln\left[ < T_t 
e^{-{i\over \hbar} \int_0^{t} dt' \int dx U(x)
\psi^\dagger(x,t)\psi(x,t)}>_{N\!S}\right] \nonumber\\
 &=&{1\over
\hbar^2}\int_0^t dt'\int_0^t dt^{\prime\prime}
K(t'-t^{\prime\prime}) \label{phase}
\end{eqnarray}
The kernel $K$ is computed assuming that the quantum dot 
does not perturb the NS point contact:
\begin{eqnarray}
K(t)={1\over 2} &&\int dx_1\int dx_2 
U(x_1)U(x_2)
\nonumber \\
&&
\times \langle\langle  \rho(x_1,t)\rho(x_2,0)+
\rho(x_2,0)\rho(x_1,t)\rangle\rangle \nonumber \\
\end{eqnarray}
This expression connects the decoherence rate Eq. (\ref{coherence rate})
 to the density fluctuations in the superconducting SPC. 

The long time behavior of the non oscillatory 
time dependence in Eq. (\ref{dot green3})
is obtained by considering the density--density kernel. 
The kernel  is characterized by a time scale $t_c$ which
identifies for which times the correlations still survive.
Therefore, in  the limit $t\rightarrow\infty$ the second
integral  over $t^{\prime\prime}$ will
saturate to a constant :  
\begin{eqnarray}
\phi(t) &\simeq& {1\over 2\hbar^2}\int_0^t dt'
\int_{-\infty}^{t'}
dt^{\prime\prime} K(t^{\prime\prime})\nonumber\\
&\simeq& {t\over\tau_\phi}~,
\end{eqnarray}
assuming that for long times, one can replace the second 
integral over the whole time domain. 

\section{Finite temperatures}
\label{appendix finite temp}

In this appendix, the energy integrals which enter in Eq. 
(\ref{rho rho distributions}) are computed analytically.

In the Andreev regime , the elements of the scattering matrix
are assumed to be weakly dependent on $\epsilon$, so one only
needs to compute the integrals  of the Fermi Dirac
distributions, which have the general form: 
\begin{eqnarray}
 \int_0^{+\infty} d\epsilon
\frac{1}{1+e^{\beta(\epsilon\mp eV)}}&&
\frac{1}{1+e^{-\beta(\epsilon-\omega \pm eV)}} =
\nonumber\\
&&\frac{-1}{\beta (e^{\beta(\omega \mp 2eV)}- 1)}
ln\Bigl\{ \frac{1+e^{\pm\beta eV}}{1+e^{\beta(\omega \mp eV)}}
\Bigr\} \nonumber\\
\end{eqnarray} 

\begin{eqnarray}
 \int_0^{+\infty} d\epsilon
\frac{1}{1+e^{\beta(\epsilon\mp eV)}}&&
\frac{1}{1+e^{\beta(-\epsilon+\omega \pm eV)}} =
\nonumber\\
&&\frac{-1}{\beta (e^{\beta \omega}- 1)}
ln\Bigl\{ \frac{1+e^{\pm \beta eV}}{1+e^{\beta(\omega \pm eV)}}
\Bigr\} \nonumber\\
\end{eqnarray}

Combining the thermal factors, the density--density correlator
can be computed as :
\begin{eqnarray}
\nonumber
&&
\langle \langle \rho(x_1,\omega)\rho(x_2,-\omega) \rangle \rangle =
\frac{2 e^2 \vert s_{eh}\vert ^2 }{h^2 \hbar^2 v_F^2} 
\nonumber
 \\
&&
\Big\{
\frac{-4}{\beta (e^{\beta(\omega -2eV)}- 1)}
ln\Bigl\{ \frac{1+e^{\beta eV}}{1+e^{\beta(\omega - eV)}}
\Bigr\}\nonumber  
\\
&& \hspace{1cm} \times 
\lbrack \vert s_{ee}\vert^{2}+s_{ee}^{*}e^{2ik_{F}x_{2}}
+s_{ee}e^{-2i k_{F}x_{1}}+e^{2ik_{F}(x_{2}-x_{1})}\rbrack\nonumber
\\
&& \hspace{0.5cm}
+
\frac{4}{\beta (e^{\beta(\omega +2eV)}- 1)}
ln\Bigl\{ \frac{1+e^{-\beta eV}}{1+e^{\beta(\omega + eV)}}\nonumber
\Bigr\} 
\\
&& \hspace{1cm} \times
\lbrack \vert s_{ee}\vert^{2}+s_{ee}^{*}e^{2ik_{F}x_{1}}
+s_{ee}e^{-2i k_{F}x_{2}}+e^{2ik_{F}(x_{1}-x_{2})}\rbrack\nonumber
\\ 
&& \hspace{0.5cm}
+ 
\frac{1}{\beta (e^{\beta \omega}- 1)}
ln\Bigl\{ \frac{1+e^{\beta eV}}{1+e^{\beta(\omega + eV)}}
\Bigr\} \nonumber
 \\  && \hspace{1cm} \times
 \lbrace 1 - (\vert s_{eh}\vert ^{2} - s_{ee} s_{hh}) + 
s_{ee} e^{-2 i k_{F} x_2 } + s_{hh} e^{2 i k_{F} x_2 } \rbrace\nonumber
\\ 
&& \hspace{0.5cm}
+
\frac{-1}{\beta \vert s_{eh}\vert ^2 (e^{\beta \omega}- 1)}
ln\Bigl\{ \frac{1+e^{-\beta eV}}{1+e^{\beta(\omega - eV)}}
\Bigr\} 
  \nonumber \\  && \hspace{1cm} \times
\Big[ 2 \vert s_{ee}\vert ^{2} \lbrace \vert s_{ee}\vert ^2 + 1 +
s_{ee}(e^{-2 i k_F x_1 } + 1/ 2 e^{-2 i k_F x_2 })
\nonumber \\ && \hspace{1.2cm} +
s_{ee}^{*} ({e^{2 i k_F x_1 }+1/2  e^{2 i k_F x_2 }})
+cos(2 k_F(x_2-x_1))\rbrace 
\nonumber
 \\
&& \hspace{1.2cm} + s_{ee}^{*} e^{2 i k_{F} x_2 }(1 + s_{ee}^{*} e^{2 i k_{F}
x_1} ) 
\nonumber
 \\
&& \hspace{1.2cm}
+ s_{ee} e^{-2 i k_{F} x_2 }(1 + s_{ee} e^{-2 i k_{F} x_1} ) \Big]\Big\}
\nonumber \\
\end{eqnarray}

At zero temperature this result corresponds to Eq. 
(\ref{rho rho zero}).


\end{multicols} 
\end{document}